\documentclass[sigconf, nonacm=true]{acmart}

\usepackage{multirow}
\usepackage{subfigure}
\usepackage{url}

\renewcommand\footnotetextcopyrightpermission[1]{} 
\AtBeginDocument{%
  \providecommand\BibTeX{{%
    \normalfont B\kern-0.5em{\scshape i\kern-0.25em b}\kern-0.8em\TeX}}}

\setcopyright{acmcopyright}
\copyrightyear{2018}
\acmYear{2018}
\acmDOI{10.1145/1122445.1122456}

\acmConference[Woodstock '18]{Woodstock '18: ACM Symposium on Neural
  Gaze Detection}{June 03--05, 2018}{Woodstock, NY}
\acmBooktitle{Woodstock '18: ACM Symposium on Neural Gaze Detection,
  June 03--05, 2018, Woodstock, NY}
\acmPrice{15.00}
\acmISBN{978-1-4503-XXXX-X/18/06}

\begin{document}

\title[]{Speaker Adaption with Intuitive Prosodic Features for Statistical Parametric Speech Synthesis}

\author{Pengyu Cheng}
\affiliation{%
	\institution{University of Science and Technology of China}
	\streetaddress{443 Huangshan Rd}
	\city{Hefei}
	\country{China}}
\email{cpy122@mail.ustc.edu.cn}

\author{Zhenhua Ling}
\affiliation{%
  \institution{University of Science and Technology of China}
  \streetaddress{443 Huangshan Rd}
  \city{Hefei}
  \country{China}}
\email{zhling@ustc.edu.cn}

\renewcommand{\shortauthors}{}

\begin{abstract}
In this paper, we propose a method of speaker adaption with intuitive prosodic features for statistical parametric speech synthesis.
The intuitive prosodic features employed in this method include pitch, pitch range, speech rate and energy considering that they are directly related with the overall prosodic characteristics of different speakers. 
The intuitive prosodic features are extracted at utterance-level or speaker-level, and are further integrated into the existing speaker-encoding-based and speaker-embedding-based adaptation frameworks respectively. The acoustic models are sequence-to-sequence ones based on Tacotron2. Intuitive prosodic features are concatenated with text encoder outputs and speaker vectors for decoding acoustic features.
Experimental results have demonstrated that our proposed methods can achieve better objective and subjective performance than the baseline methods without intuitive prosodic features.   
Besides, the proposed speaker adaption method with utterance-level prosodic features has achieved the best similarity of synthetic speech among all compared methods.
\end{abstract}




\maketitle

\section{Introduction}
Recently, with the development of deep learning techniques, text-to-speech (TTS) systems have been able to synthesize high quality voice 
and have been applied in many areas, such as simultaneous interpretation, news reading and virtual characters, etc.
Statistical parametric speech synthesis (SPSS) with sequence-to-sequence (seq2seq) acoustic models \cite{8461368,li2019neural,ren2019fastspeech} is the state-of-the-art approach to building TTS systems nowadays.
However, these models usually rely on a large amount of audio samples from one single speaker (e.g., 10 to 20 hours) for high quality speech synthesis. This severely limits the effectiveness of above models in low resource situations. 
Therefore, the task of speaker adaptation, or voice cloning, which transfers a pre-trained multi-speaker acoustic model to a specific target speaker with limited data has attracted much research attention recently. Existing studies on this task can be divided into two main categories, i.e., \emph{speaker-encoding}-based ones \cite{speaker_encoding_1, nachmani2018fitting, hu2019neural, kons2019high,cooper2020zero, jia2018transfer} and \emph{speaker-embedding}-based ones \cite{arik2018neural,chen2018sample, speaker_embedding_adaption, deng2018modeling}.

In the speaker-encoding-based methods \cite{speaker_encoding_1, nachmani2018fitting, hu2019neural, kons2019high}, a neural speaker encoder is employed to provide a vector describing speaker characteristics for each utterance, 
and this speaker vector is further combined with the outputs of text encoder for decoding acoustic features.
Such speaker encoders are usually 
learned by a speaker verification task on a large-scale multi-speaker dataset and are fixed when training  synthesis models.
For a specific target speaker, speaker adaption is conducted by fine-tuning model parameters with limited data from the target speaker or just calculating a new speaker vector using a reference recording of the target speaker without fine-tuning any model parameters 
\cite{cooper2020zero, jia2018transfer}. 
The latter can achieve zero-shot adaptation for unseen speakers.

In the speaker-embedding-based methods \cite{arik2018neural,chen2018sample, speaker_embedding_adaption, deng2018modeling}, a speaker embedding vector is utilized to represent the overall characteristics of each speaker.
This embedding vector is also combined with the outputs of text encoder to predict acoustic features.
At the stage of pre-training with a multi-speaker dataset, the embedding vectors of all speakers are learned from scratch together with other model parameters. 
For a specific target speaker, a new embedding vector is assigned to this speaker and is estimated using the adaptation data.

On the other hand, prosody characteristics influence the subjective perception of synthetic speech significantly. Recently, there have been many studies on prosody modeling in TTS. 
Google proposed an unsupervised prosody learning method \cite{generalstyletokens} for single speaker. Similar to \cite{generalstyletokens}, variational autoencoder (VAE) was also introduced to learn the latent representation of speaking styles in an unsupervised manner \cite{prosody_VAE}. These methods can model prosodic features in a latent representation space and need no extra labels of prosody information. However, the learned latent representation space usually contains all residual acoustic information that can not be described by the linguistic input, e.g., background noise, instead of focusing on prosodic variations. 
Some other methods introduced intuitive prosodic features to enhance the performance of TTS models. Fastspeech2 \cite{fastspeech2} utilized prosodic features including pitch, energy and duration at phoneme level. Similarly, Apple \cite{apple_prosody_control} proposed to utilize utterance-level intuitive prosodic features to control the style of synthetic speech from different prosodic aspects. 

Considering that prosody characteristics are also important for discriminating different speakers, some methods integrating prosody information into multi-speaker modeling and speaker adaption have also been proposed. Attentron \cite{attentron} introduced a fine-grained prosody encoder that focused on frame-level prosody information. Adaspeech \cite{chen2021adaspeech} made full use of the information at multiple levels of speech for acoustic model training. 
However, all these methods employed latent prosodic representations for acoustic modeling and no studies on utilizing intuitive prosodic features for speaker adaptation have been conducted.

Therefore, this paper proposes a method of multi-speaker training and speaker adaptation with intuitive prosodic features for seq2seq speech synthesis. 
Similar to the study on controllable speech synthesis with intuitive prosodic features \cite{apple_prosody_control}, the intuitive prosodic features used in this method consist of four variables, 1) pitch, 2) pitch range, 3) speech rate, and 4) energy.
Different from previous studies \cite{generalstyletokens, attentron} which focused on short-time prosodic features, we calculate intuitive prosodic features  at utterance-level or speaker-level in this paper.
Comparing with short-time prosodic features and latent prosodic representations, the advantage of long-term intuitive prosodic features is that they are explicitly meaningful and are directly related with the overall rhythmic properties of different speakers. 
Then, two Tacotron2-based model architectures are designed that incorporate utterance-level intuitive prosodic features and speaker-level intuitive prosodic features into the existing speaker-encoding-based and speaker-embedding-based speaker adaption frameworks respectively.
In both architectures, the intuitive prosodic features are combined with the outputs of the text encoder for decoding acoustic features.
The model parameters are first pre-trained using a multi-speaker dataset and are then fine-tuned on the adaptation data of the target speaker.
Experimental results on the AISHELL-3 dataset \cite{aishell3_dataset} demonstrate that integrating  intuitive prosodic features effectively improved the objective distortion and subjective similarity of speaker adaptation under both speaker-encoding-based and speaker-embedding-based frameworks. Besides, the proposed speaker adaptation method with utterance-level intuitive prosodic features achieved the best similarity score of converted speech among all compared methods\footnote{Speech samples can be found at  \href{http://home.ustc.edu.cn/~cpy122/SAIPF/}{http://home.ustc.edu.cn/$\sim$cpy122/SAIPF/}.}.

\begin{figure}[t]
	\centering
	\includegraphics[width=8.5cm]{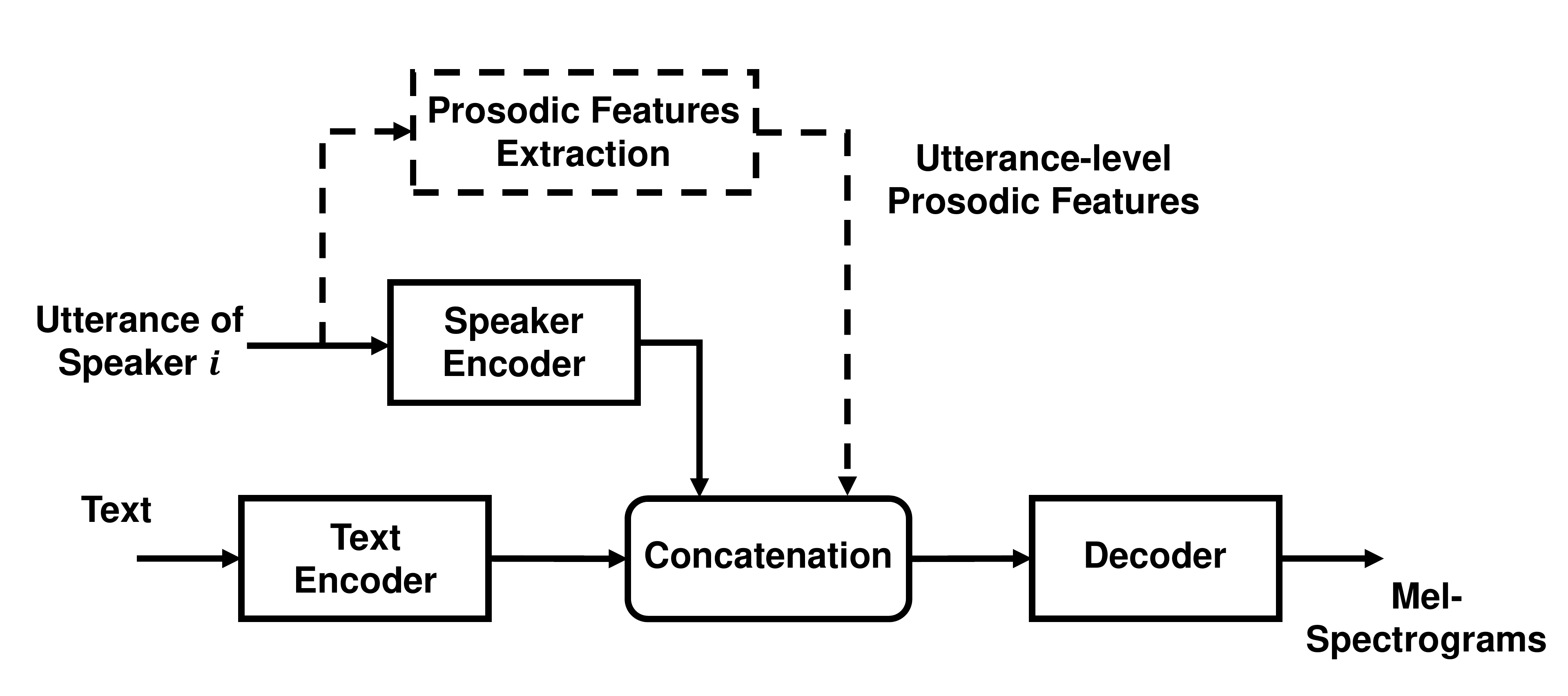}
	\caption{The model architecture of speaker adaption with utterance-level prosodic features. The dotted lines indicate our modifications to the \emph{speaker-encoding}-based adaptation framework.}
	\label{fig:utterance-level model structure}
\end{figure}

\section{Methods}
This paper designs two speaker adaption models on the basis of Tacotron2 \cite{8461368}.
These two models follow the speaker-encoding-based and the speaker-embedding-based speaker adaptation frameworks respectively. 
In addition to the speaker vectors given by the speaker encoder or the embedding table, we introduce intuitive prosodic features extracted from audio to better describe perception-related prosodic characteristics of different speakers. 
Details of our proposed methods are described in this section.

\subsection{Intuitive Prosodic Features}
\label{prosodic features}
The intuitive prosodic features employed in this paper include fundamental frequency (pitch), pitch range, speech rate and energy. These features can be easily extracted from audio and can reflect the specific prosodic characteristics of different speakers. For example, the fundamental frequencies of male and female voices are very different. 
Elderly people usually have narrower pitch range and slower speech rate than young people.
In this paper, we extract intuitive prosodic features at utterance-level and at speaker-level respectively.

When extracting utterance-level prosodic features, a pitch estimator is first employed to extract frame-wise fundamental frequencies ($F_0$) of utterances. 
Then, the utterance-level pitch is calculated through averaging the log-$F_0$ of all voiced frames in an utterance.
The utterance-level pitch range is defined as the difference between the highest log-$F_0$ and the lowest log-$F_0$ in an utterance.
Considering the possible errors made by the pitch estimator, we ignore the frames with the highest $5\%$ and  the lowest $5\%$ log-$F_0$ in an utterance when calculating the pitch range.
Given the phoneme transcription of an utterance, we use the MFA \cite{montreal_force_aligner} toolkit for forced alignment and calculate the average duration of all phonemes within a sentence to get the utterance-level speech rate.  
Finally, we calculate frame-wise energies from waveforms after silence trimming 
and the utterance-level energy is computed by averaging all frame-wise energies on the non-silence segments in an utterance.

Regarding with speaker-level prosodic features, we simply average above utterance-level features of all utterances from one speaker to obtain the speaker-level ones.

\begin{figure}[t]
	\centering
	\includegraphics[width=8.5cm]{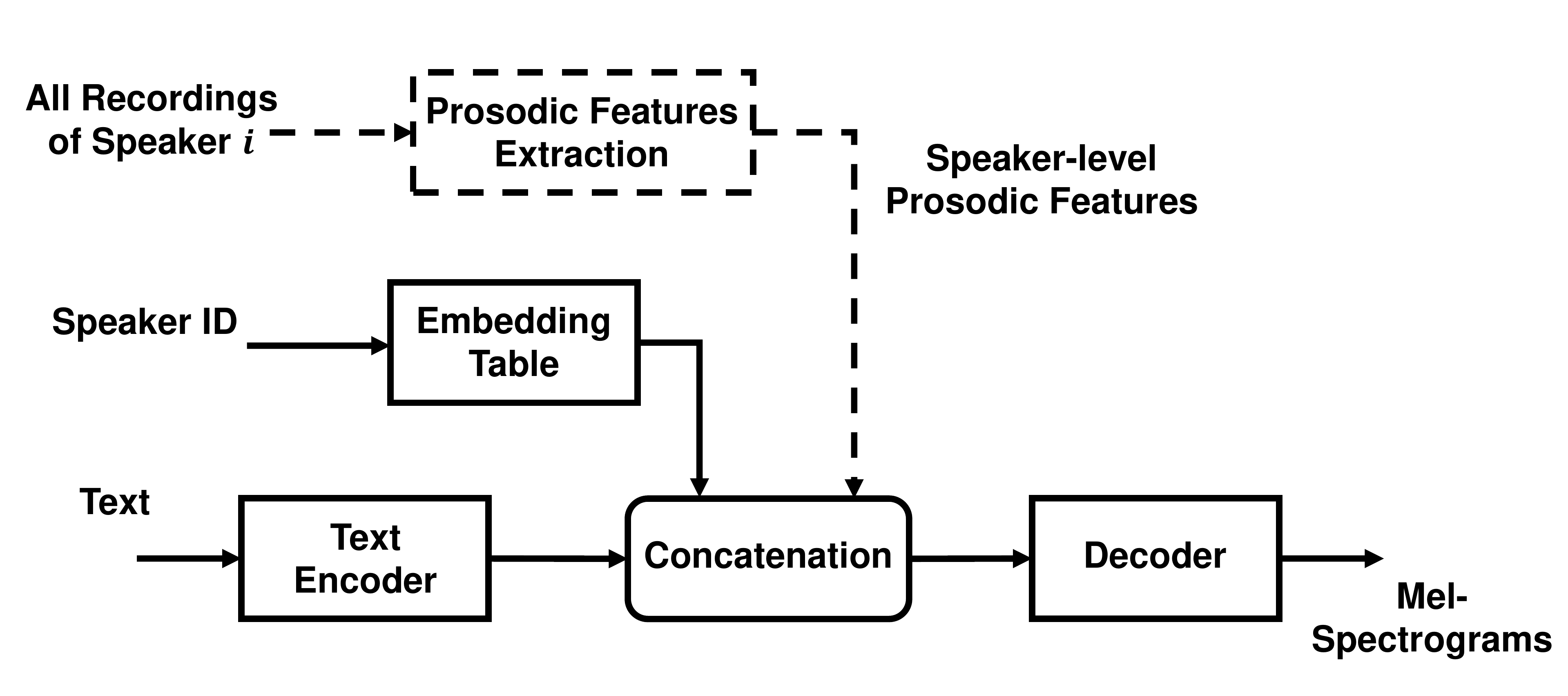}
	\caption{The model architecture of speaker adaption with speaker-level prosodic features. The dotted lines indicate our modifications to the \emph{speaker-embedding}-based adaptation framework.}
	\label{fig:speaker-level model structure}
\end{figure}

\subsection{Speaker Adaption with Utterance-Level Prosodic Features}
\label{utterence-level-architecture}
The model architecture of our proposed speaker adaption method with utterance-level prosodic features is shown in Fig. \ref{fig:utterance-level model structure}.
This method follows the speaker-encoding-based adaptation framework  and 
consists of four parts, a prosodic feature extraction module, a speaker encoder, a text encoder and a decoder.
Here, the prosodic feature extraction module extracts utterance-level intuitive prosodic features from an utterance, as we have mentioned in Section \ref{prosodic features}.
We follow the structure in \cite{utterance-speaker-embedding} to build the speaker encoder, 
whose inputs are Mel-spectrograms of an utterance with variable length. 
The input vector of its softmax layer is defined as the speaker vector, which has a fixed dimension.
The text encoder maps input linguistic feature sequences into hidden vector sequences. 
After concatenating the outputs of the prosodic feature extraction module, the speaker encoder, and the text encoder, the decoder generates Mel-spectrograms frame-by-frame in an autoregressive way.
Both the text encoder and the decoder are similar to the ones in Tacotron2 \cite{8461368}.

Let $\textbf{X}$ denote the  linguistic feature sequence of input text with length $N$.
The text encoder output $\textbf{E}_{out} \in \mathbb{R}^{d_{e} \times N} $ can be calculated as
\begin{align}
	\textbf{E}_{out} = Encoder(\textbf{X}),
\end{align}
where $d_{e}$ is the feature dimension of encoder outputs. 
Let $\textbf{s}\in \mathbb{R}^{d_s}$ and $\textbf{p}\in \mathbb{R}^{d_p}$ denote the output vectors of the speaker encoder and the prosodic feature extraction module, respectively.
Here, $d_p=4$ as introduced in Section \ref{prosodic features}.
Then, we tile $\textbf{s}$ and $\textbf{p}$ along time axis by repetition to match the length of $\textbf{E}_{out}$, and get $\textbf{S}_{tiled} \in \mathbb{R}^{d_{s} \times N}$ and
$\textbf{P}_{tiled} \in \mathbb{R}^{ d_{p} \times N}$.
Finally, the decoding process can be written as
\begin{align}
	\textbf{D}_{in} = [\textbf{E}_{out}; \textbf{S}_{tiled}; \textbf{P}_{tiled}],
\end{align}
\begin{align}
	\textbf{O} = Decoder(\textbf{D}_{in}),
\end{align}
where 
$\textbf{D}_{in} \in \mathbb{R}^{(d_{e}+d_{s}+d_{p}) \times  N}$ and $\textbf{O}$ represents the decoded Mel-spectrograms corresponding to the input text.

The speaker encoder is learned in advance via a speaker verification task and its parameters are fixed when training other modules. The text encoder and the decoder are learned by pre-training using a multi-speaker dataset and we only fine-tune the decoder using the adaptation data of a specific target speaker. 
At these two stages, both speaker vectors and prosodic features are utterance-level ones. 
The loss function contains the mean squared error (MSE) between generated Mel-Spectrograms and ground truth ones together with a stop token loss, just like Tacotron2 \cite{8461368}.

At the inference stage, since the audio corresponding to the input text to be synthesized is not available,
we average all utterance-level speaker vectors and prosodic features of adaptation data, and concatenate them with the text encoder outputs for decoding.
In our preliminary experiments, this strategy achieved better performance than choosing one single reference utterance from adaptation data. 

\subsection{Speaker Adaption with Speaker-Level Prosodic Features}
\label{speaker-level-architecture}
The model architecture of our proposed speaker adaption method with speaker-level prosodic features  is illustrated in Fig. \ref{fig:speaker-level model structure}.
This method follows the speaker-embedding-based adaptation framework.
Its calculation processes are the same as the ones introduced in Section \ref{utterence-level-architecture} except two main differences. 
First, the speaker encoder in Fig. \ref{fig:utterance-level model structure} is replaced by an embedding table in Fig. \ref{fig:speaker-level model structure} which contains a trainable embedding vector for each speaker in the training set.
Second, to be compatible with the speaker-embedding-based framework, speaker-level prosodic features are employed for decoding Mel-spectrograms in Fig. \ref{fig:speaker-level model structure}.
In other words, for a specific speaker, all his (or her) utterances are represented by one same speaker  vector and same intuitive prosodic features.

At the multi-speaker pre-training stage, the embedding table is jointly optimized with other model parameters.
At the adaptation stage, a new embedding vector for the target speaker is estimated and only the decoder is fine-tuned.
The loss function is the same as the ones introduced in Section \ref{utterence-level-architecture}.
At the inference stage, speaker-level speaker vector and prosodic features are utilized for decoding, just like the training stages.

\begin{figure}[t]
	\centering
	\includegraphics[width=8.5cm]{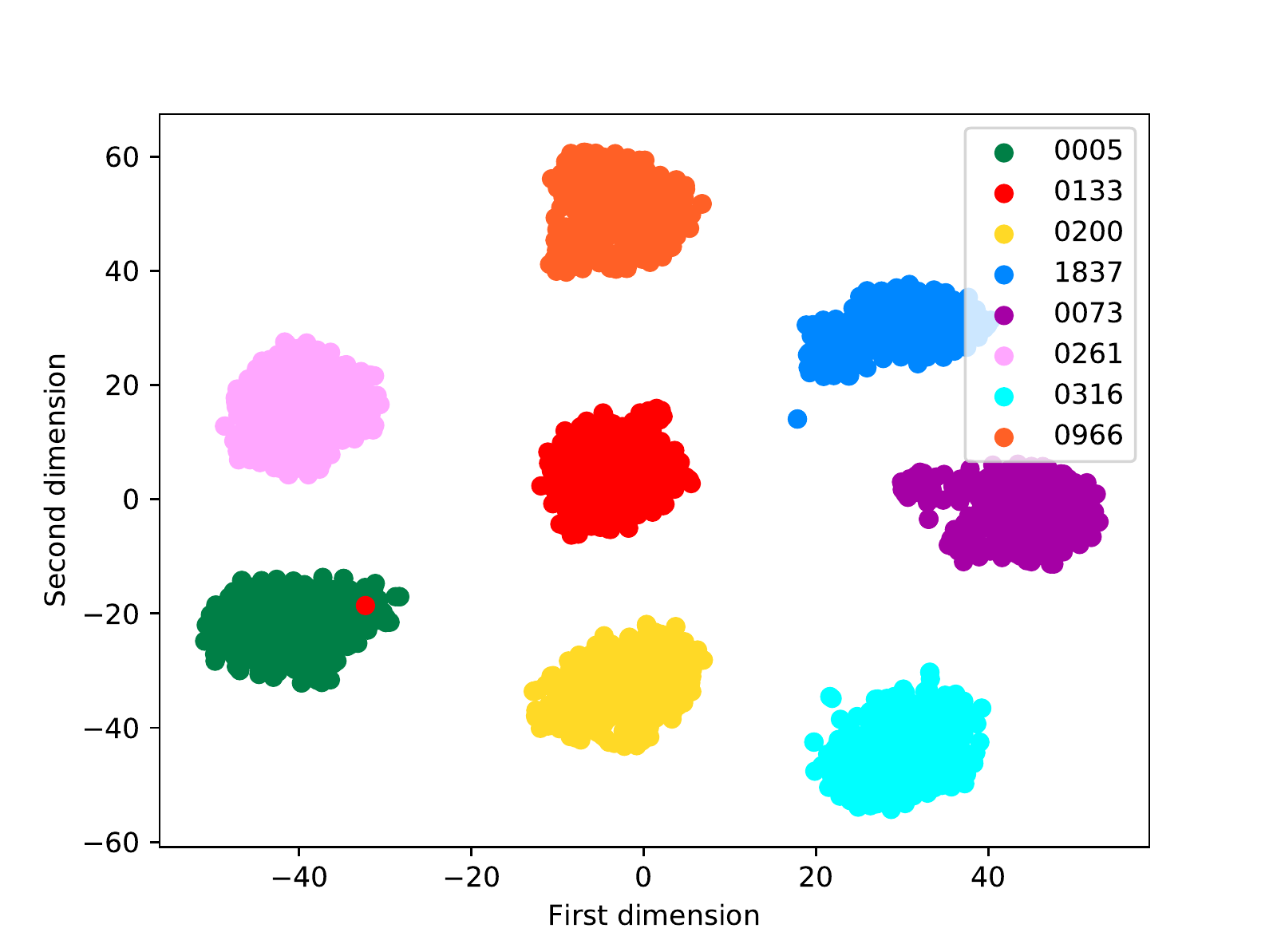}
	\caption{t-SNE visualization of utterance-level speaker vectors for the 8 unseen speakers. The numbers in the legend indicate their speaker IDs in AISHELL-3 corpus.}
	\label{fig:tSNE_result}
\end{figure}

\section{Experiments}
\label{sec:exp}
\subsection{Datasets}
\label{subset:datasets}
The AISHELL-3\cite{aishell3_dataset} Mandarin corpus was adopted in our experiments. 
The dataset contains 218 speakers with 88,035 utterances. 
We found that some linguistic feature sequences of utterances contained no information about phrase boundaries, so we discarded these samples without such information. 
The remaining set has a total of 63,623 samples and contains 174 (31 male and 143 female) speakers. 
In our experiments, 8 (4 male and 4 female) speakers were randomly chosen as unseen target speakers for adaptation. 
Furthermore, 80, 10 and 20 utterances were randomly selected from the audio samples of each target speaker to form his (or her) adaption set, validation set and test set respectively. 
The remaining 166 speakers were used as the seen multi-speaker dataset for pre-training. 
For each seen speaker, we divided all his (or her) samples into a train set and a development set with a ratio of 19:1.
This validation set was adopted to prevent over-fitting when pre-training multi-speaker  models.

When implementing  our proposed speaker adaption method with utterance-level prosodic features,
we utilized an existing speaker encoder \cite{aishell3_dataset}, which had a ResNet-based architecture and was trained on a large scale dataset containing 775,289 utterances from aidataTang \cite{aidatatang} and MAGICDATA \cite{magicdata} datasets. 
The speaker vectors of the utterances from the 8 unseen speakers are  extracted by this speaker encoder and are visualized in Fig. \ref{fig:tSNE_result}  using t-distributed stochastic neighbour embedding (t-SNE) method. 
We can see all speaker vectors are clearly clustered corresponding to speaker IDs. 


\begin{table}[t]
	\renewcommand{\arraystretch}{1.4}
	\caption{Objective evaluation results of four systems on the test sets of female and male speakers. 
	}
	\label{table_obj_eva}
	\centering
	\begin{tabular}{c|c|c|c|c}
		\hline
		\hline
		\multirow{3}{*}{Models} & \multicolumn{2}{|c|}{Female Speakers} & \multicolumn{2}{|c}{Male Speakers} \\
		\cline{2-5}
		& MCD & $F_{0}$ RMSE & MCD & $F_{0}$ RMSE \\
		& (dB) & (Hz) & (dB) & (Hz) \\
		\hline
		\emph{Baseline-Enc} & 3.8509 & 47.0806 & 5.3637 & 27.7887 \\
		\hline
		\emph{Baseline-Emb} &  4.2181 & 49.4417 & 6.2164 & 36.7553 \\
		\hline
		\emph{UPF-Enc} & \textbf{3.7626} & \textbf{44.0754} & \textbf{5.1064} &  \textbf{26.9880}\\ 
		\hline
		\emph{SPF-Emb} & 4.0447 & 47.8555 & 5.3907 & 28.5118 \\
		\hline
		\hline
	\end{tabular}
\end{table}

\subsection{Experimental Configurations}
We downsampled the audio of AISHELL-3 to 16kHz. 
Spectrograms were analyzed by 800 point short-time Fourier transform (STFT) with Hann windowing. The frame shift and length were 12.5 ms and 50 ms. 
Then, 80-dimensional Mel-spectrograms were computed from pre-emphasized speech.

At the training stage, we used the Adam optimizer with $\beta1 = 0.9$, $\beta2 = 0.999$, $\epsilon = 10^{-6}$. 
Model parameters were pre-trained on the multi-speaker dataset for 70k steps. 
The learning rate was initially set to $10^{-3}$ and was reduced to $10^{-5}$ after 50k steps. 
The batch size was 40. 
At the adaption stage,  
we fine-tuned the model for fixed 600 steps for each unseen speaker. 
The batch size was reduced to 20 and the learning rate was fixed at $10^{-5}$. 
The entire training process was done on a TITAN RTX GPU with 25G RAM. 
The pre-training stage costed about 36 hours while the adaptation stage needed about 30 minutes for each speaker.

\subsection{System Construction}
To verify the effectiveness of our proposed method, four systems were constructed for comparison. 

\emph{1) Speaker-encoding-based baseline (Baseline-Enc)}: The architecture of \emph{Baseline-Enc} is shown as the solid lines in Fig \ref{fig:utterance-level model structure}. The speaker encoder introduced in Section \ref{subset:datasets} was adopted here to extract utterance-level speaker vectors.
At  multi-speaker training and adaption stages, the parameters of the speaker encoder were fixed.
At the inference stage, the input speaker vector was the average of all speaker vectors calculated from the adaption utterances of the target speaker.

\emph{2) Speaker-embedding-based baseline (Baseline-Emb)}: Its architecture is illustrated by the solid lines in Fig \ref{fig:speaker-level model structure}.
The speaker vectors were columns in an embedding table, which were jointly optimized with other model parameters. At multi-speaker training, adaption and inference stages, the input speaker information was simply speaker ID. 

\emph{3) Proposed method with utterance-level prosodic features (UPF-Enc)}: This system introduced utterance-level prosodic features into \emph{Baseline-Enc} as illustrated in Fig \ref{fig:utterance-level model structure}.
Its implementation followed the introductions in Section \ref{utterence-level-architecture}.

\emph{4)  Proposed method with speaker-level prosodic features (SPF-Emb)}: This system introduced speaker-level prosodic features into \emph{Baseline-Emb} as illustrated in Fig \ref{fig:speaker-level model structure}.
Its implementation followed the introductions in Section \ref{speaker-level-architecture}.

In the above four systems, the detailed configurations of the text encoder and the decoder were consistent with that in Tacotron2 \cite{8461368}.
The dimensionality of speaker vectors $d_{s}$ was 256.
To speed model convergence, L2 normalization was applied to both speaker vectors and prosodic features. 

All these systems adopted the same Parallel WaveGAN \cite{yamamoto2020parallel} vocoder for reconstructing waveforms from the decoded Mel-scale spectrograms.
Since this paper focuses on acoustic modeling for speaker adaption, we simply trained the vocoder on all AISHELL-3 data without adaption for specific target speakers.

\begin{table}[t]
	\renewcommand{\arraystretch}{1.4}
	\caption{Objective evaluation results in ablation studies of proposed prosodic features. 
	}
	\label{table_obj_eva_ablation}
	\centering
	\begin{tabular}{c|c|c|c|c}
		\hline
		\hline
		\multirow{3}{*}{Models} & \multicolumn{2}{|c|}{Utterance-level} & \multicolumn{2}{|c}{Speaker-level} \\
		\cline{2-5}
		& MCD & $F_{0}$ RMSE & MCD & $F_{0}$ RMSE \\
		& (dB) & (Hz) & (dB) & (Hz) \\
		\hline
		proposed & \textbf{4.2383} & \textbf{35.7716} & \textbf{4.7032} & \textbf{38.5562} \\
		\hline
		\emph{-pitch} &  4.5928 & 37.1103 & 5.1231 & 41.3770 \\
		\hline
		\emph{-pitch range} & 4.3301 & 36.1094 & 4.9362 & 39.5105 \\ 
		\hline
		\emph{-speech rate} & 4.5473 & 36.7246 & 5.2498 & 40.2838 \\
		\hline
		\emph{-energy} & 4.3729 & 36.2938 & 4.8232 & 39.5903 \\
		\hline
        \emph{-all} & 4.6629 & 37.0948 & 5.2148 & 42.7793 \\
		\hline
		\hline
	\end{tabular}
\end{table}

\begin{figure*}[t]
	\centering
	\subfigure[Mean opinion scores of male speaker \textbf{0966}.]{
		\includegraphics[width=8cm]{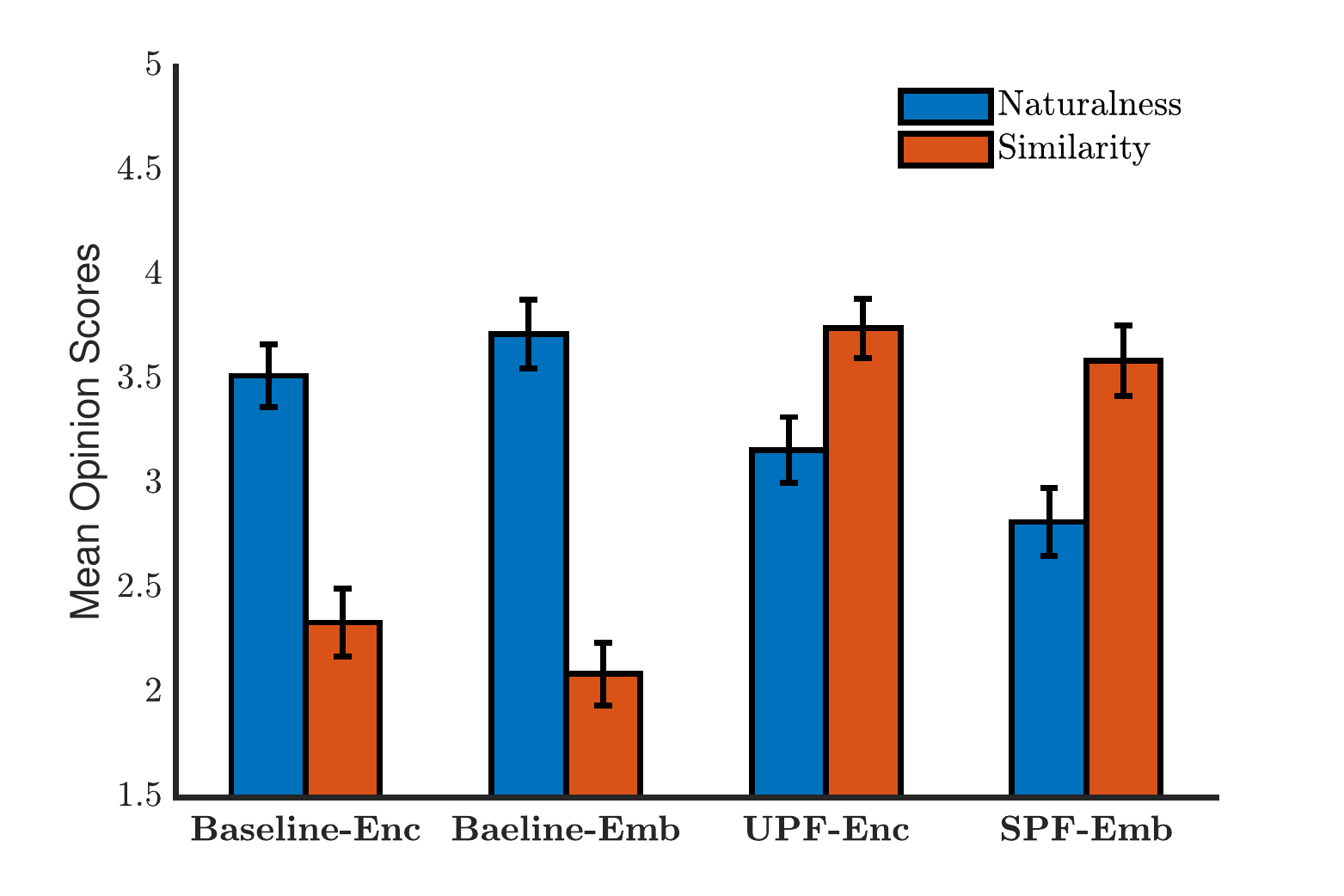}
		\label{0966}
	}
	\quad
	\subfigure[Mean opinion scores of female speaker \textbf{0133}.]{
		\includegraphics[width=8cm]{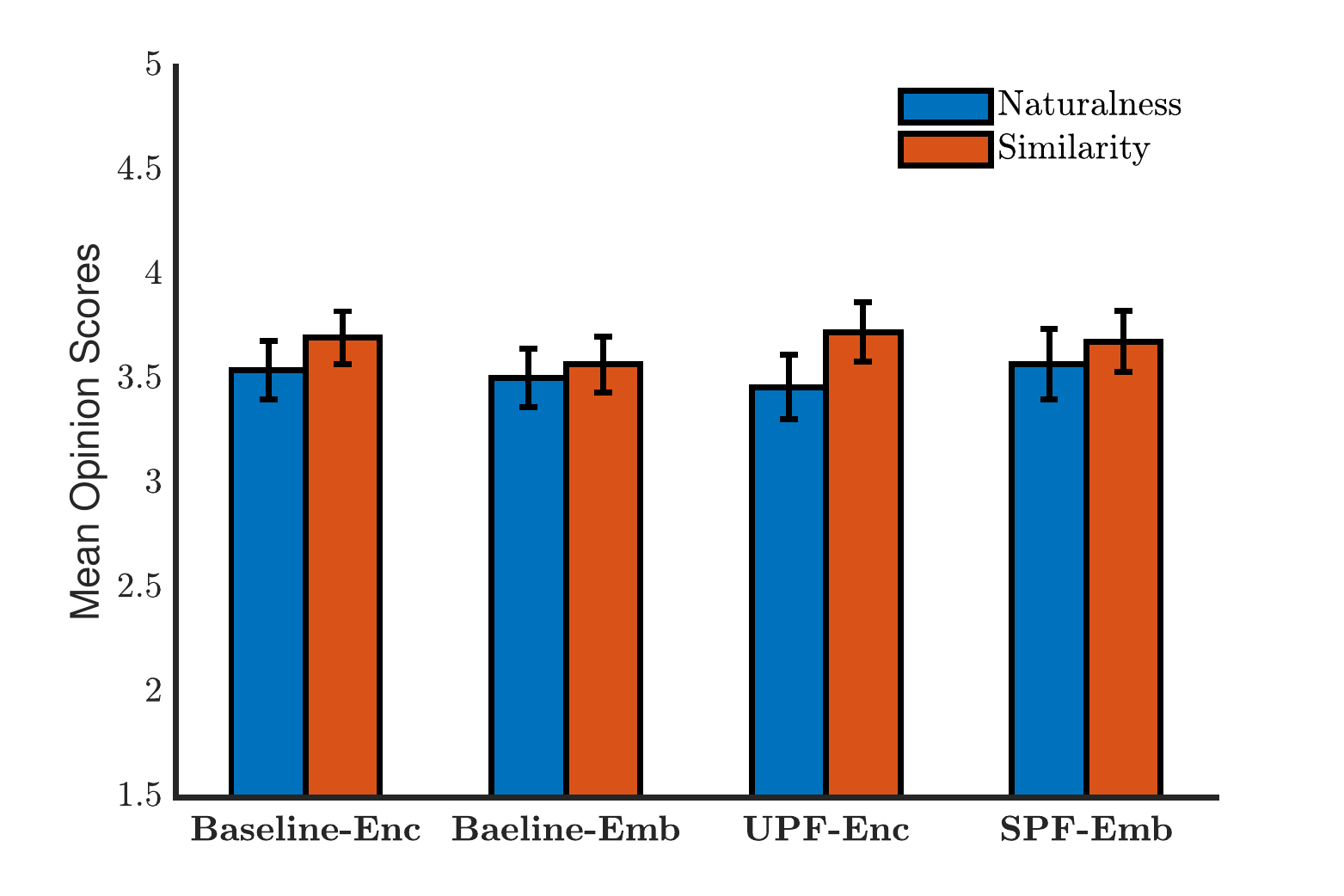}
		\label{0133}
	}
	\quad
	\subfigure[Mean opinion scores of male speaker \textbf{0073}.]{
		\includegraphics[width=8cm]{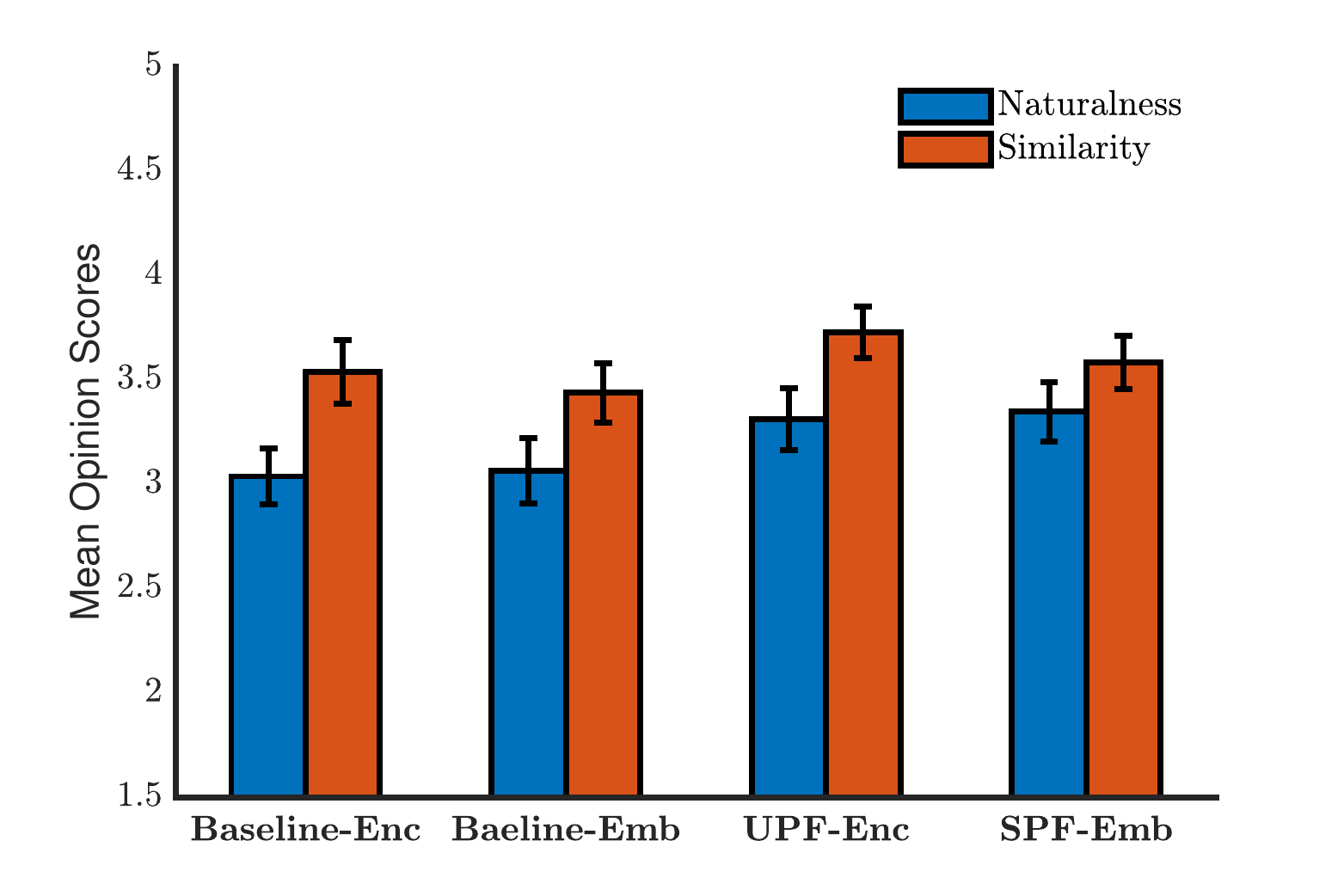}
		\label{0073}
	}
	\quad
	\subfigure[Mean opinion scores of female speaker \textbf{0005}.]{
		\includegraphics[width=8cm]{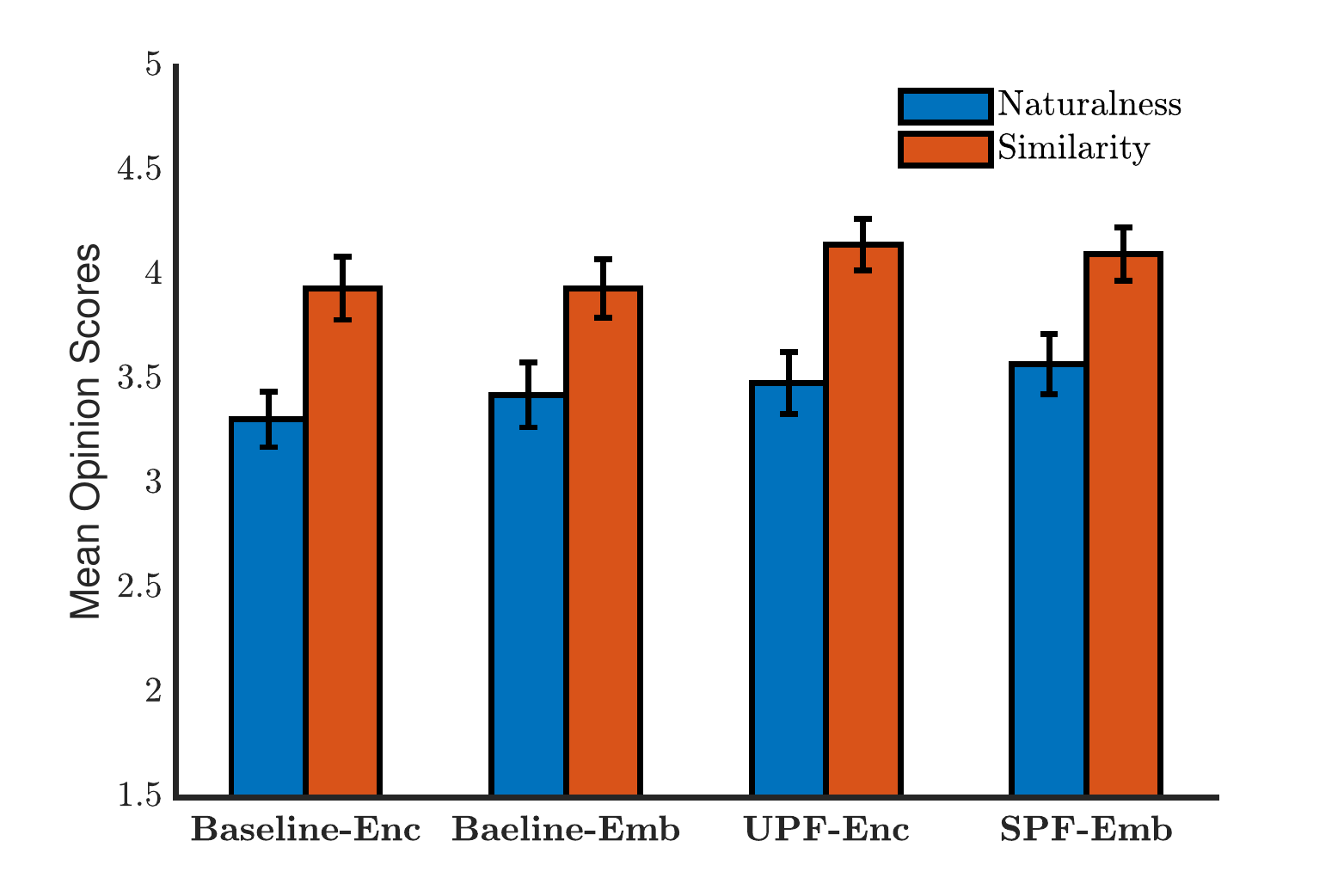}
		\label{0005}
	}
	\caption{Mean opinion scores with $ 95\% $ confidence intervals of different systems on test set of four speakers. Higher is better.}
	\label{MOS results}
\end{figure*}

\subsection{Objective Evaluation Results}
\label{obj eva res}
For objective evaluation, $F_{0}$ and 25-dimensional Mel-cepstral coefficients (MCCs)  were extracted from reconstructed waveforms. 
Mel-cepstrum distortion (MCD) and root mean square error of $F_{0}$ ($F_{0}$ RMSE)  were adopted as objective evaluation metrics.
Since the synthetic speech and ground-truth recordings were not aligned, dynamic time warping (DTW) \cite{mcd_dtw} was performed based on MCCs before calculating MCD and $F_{0}$ RMSE.

The evaluation results on the test sets of female and male speakers are reported in Table \ref{table_obj_eva}.
We can see that the MCDs of male speakers were much higher than that of females speakers for all four systems.
One possible reason is the imbalanced gender distribution in AISHELL-3 dataset.
In the original AISHELL-3 dataset, the ratio between female and male speakers is 175:43.
While in our multi-speaker dataset for pre-training, the ratio is 140:26.
Such imbalance increased the difficulty of adaptation for male target speakers.
On the other hand, the $F_{0}$ RMSEs of male speakers were much smaller than that of female speakers due to the inherently lower pitch and narrower pitch range of male speakers.  

Comparing the two baseline systems,  \textit{Baseline-Enc} performed better than \textit{Baseline-Emd}.
Comparing \textit{Baseline-Enc} with \textit{UPF-Enc} and \textit{Baseline-Emb} with \textit{SPF-Emb}, we can see that after introducing intuitive prosodic features, the MCDs and $F_{0}$ RMSEs  of both baseline systems were reduced.
This demonstrates that our proposed methods can effectively decrease the distortion between the predicted acoustic features and natural ones.
Furthermore, \textit{UPF-Enc} achieved the best objective performance among the four systems for both female and male speakers.

In order to analyse the effects of different prosodic features in our proposed method, ablation experiments were further conducted.
For investigating the effect of specific prosodic feature, we removed this feature from model (e.g., \textit{"-pitch"} ).
Thus, four different ablation systems were constructed.
In addition, \textit{"-all"} means that all four prosodic features were removed from the model, which was equivalent to baseline models (\textit{Baseline-Enc} and \textit{Baseline-Emb}).
All these systems shared the same training strategy and the architecture except for employment of prosodic features.
Corresponding to the previous experiments, all the above ablation studies were conducted separately under two systems, utterance-level and speaker-level.

Table \ref{table_obj_eva_ablation} shows the results of ablation experiments.
We can see that the performances of four ablation systems all degraded, which demonstrates the effectiveness of using the four prosodic features in our proposed method.
Moreover, all four ablation systems performed better than baseline (\textit{"-all"}), also demonstrating the validity of each feature.
It can be found that the performance of proposed method degraded drastically without using pitch and speech rate while the results became less worse without using pitch range and energy.
It may indicate that the former two prosodic features (pitch and speech rate) contribute more to the performance improvement than the other two features.

\subsection{Subjective Evaluation Results}
In order to control the scale of listening tests, 
our subjective evaluation randomly selected four (2 female and 2 male)  speakers from the 8 target speakers. 
Four listening tests were conducted for these four target speakers separately. 
For each speaker, 10 utterances in the test set were randomly chosen and were synthesized using the four systems respectively. 
At least 11 native listeners were involved in each test.
They were asked to give a 5-scale opinion score 
on both naturalness and similarity for each synthetic utterance.
Fig. \ref{MOS results} shows the mean opinion score (MOS) results on naturalness and similarity for the four target speakers.
Paired t-tests were also conducted to investigate the significance of MOS differences between different systems.

Comparing the MOS results of male speakers (Fig \ref{0966} and \ref{0073}) with female speakers (Fig \ref{0133} and \ref{0005}), 
we can see that the overall scores (both naturalness and similarity) of female speakers were higher than that of male speakers, 
which may also be attributed to the gender-imbalance issue in the AISHELL-3 dataset, as mentioned above in Section \ref{obj eva res}.
Comparing \textit{Baseline-Enc} and \textit{Baseline-Emb}, \textit{Baseline-Enc} achieved higher similarity MOS than \textit{Baseline-Emb} for speaker 0966 ($ p=7.0\times10^{-5} $) and the similarity differences were insignificant for the other three speakers ($ p>0.05 $).  
Regarding with naturalness, \textit{Baseline-Emb} outperformed \textit{Baseline-En}c ($ p=0.018 $) for speaker 0966 and their differences were insignificant for the other three speakers ($ p>0.05 $). 

Comparing \textit{UPF-Enc} and \textit{Baseline-Enc}, we can see that our proposed \textit{UPF-Enc} obtained higher similarity MOS for all four target speakers.
The differences were significant ($ p < 0.05 $) for three of them ($ p = 1.79\times 10^{-26} $ for speaker 0966, $ p = 0.04 $ for speaker 0073 and $ p = 0.008 $ for speaker 0005). 
These results demonstrate that introducing utterance-level intuitive prosodic features into speaker-encoding-based adaption framework contributes to generate audio that is more similar to target voice.
For naturalness MOS results, \textit{UPF-Enc} outperformed \textit{Baseline-Enc} for speaker 0073 ($ p = 0.001 $) and 0005 ($ p = 0.02 $) with significant differences, while the latter achieved significantly better naturalness for speaker 0966 ($ p = 0.0003 $).
One possible reasons is that speaker 0966 has a very low pitch. Although \textit{UPF-Enc} greatly increased the similarity of \textit{Baseline-Enc}, low pitch made the synthetic speech sound unnatural.

Similarly, it can be found that our proposed \textit{SPF-Emb} achieved higher similarity MOS than \textit{Baseline-Emb} for all four speakers, among which two speakers  had significant differences ($ p = 1.42\times10^{-26} $ for speaker 0966 and $ p = 0.007 $ for speaker 0005).
These results indicate the effectiveness of using speaker-level intuitive prosodic features.
As for naturalness MOS results, \textit{SPF-Emb} performed better than \textit{Baseline-Emb} for three speaker (0133, 0073 and 0005). 
However, two of these differences were insignificant ($ p > 0.05$) except for speaker 0073 ($ p = 0.001 $).

Furthermore, comparing \textit{UPF-Enc} and \textit{SPF-Emb}, we can see that the former achieved higher similarity MOS  for four speakers. 
The differences for two of them (speaker 0966 and 0073) were significant ( $ p = 0.006 $ and $ 0.047 $ respectively).
For naturalness, \textit{SPF-Emb} outperformed \textit{UPF-Enc} on speaker 0966 ( $ p = 1.013\times10^{-5} $) and the MOS differences were insignificant for the other there speakers ($ p>0.05 $). 
In summary, \textit{UPF-Enc} achieved the best similarity performance among all four systems, which is consistent with our previous objective evaluation results.

\section{Conclusions}
In this paper, we have proposed a speaker adaption architecture with intuitive prosodic features.
To demonstrate the effectiveness of our proposed model, we built two models with prosodic features of different level, Utterance-level Prosodic Features with Speaker Encoding (\textit{UPF-Enc}) and Speaker-level Prosodic Features with Speaker Embedding (\textit{SPF-Emb}).
In addition, two baseline model were constructed according to different speaker representation level, Baseline with Speaker Encoding (\textit{Baseline-Enc}) and Baseline with Speaker Embedding (\textit{Baseline-Emb}).
Experimental results have demonstrated that our proposed models with prosodic features can achieve lower acoustic distortion and higher subjective similarity 
compared to baseline models.
Our future work will focus on integrating more intuitive acoustic features that can describe speaker characteristics into our model and investigating the disentanglement between intuitive prosodic features and the speaker vectors given by speaker encoding or embedding.

\section*{acknowledgements}
This work was supported in part by the National Key R\&D Program of China under Grant 2019YFF0303001, and in part by the National Nature Science Foundation of China under Grant 61871358.

\bibliography{apsipa}
\bibliographystyle{abbrv}

\end{document}